\documentclass[conference]{IEEEtran}
\IEEEoverridecommandlockouts
\usepackage{cite}
\usepackage{amsmath,amssymb,amsfonts}
\usepackage{url}
\usepackage{algorithmic}
\usepackage{graphicx}
\usepackage{textcomp}
\usepackage{xcolor}
\usepackage{enumitem} 
\usepackage{multirow}
\usepackage{booktabs}
\usepackage{adjustbox}
\usepackage{float} 
\usepackage[tableposition=bottom]{caption} 

\def\BibTeX{{\rm B\kern-.05em{\sc i\kern-.025em b}\kern-.08em
    T\kern-.1667em\lower.7ex\hbox{E}\kern-.125emX}}
\begin{document}

% \title{MNT-Bench\\
% {\footnotesize \textsuperscript{*}Note: Sub-titles are not captured for https://ieeexplore.ieee.org  and
% should not be used}
% %\thanks{Identify applicable funding agency here. If none, delete this.}
% }

\title{MTT-Bench: Predicting Social Dominance in Mice via Multimodal Large Language Models}
\author{
Yunquan Chen$^{1}$,
Haoyu Chen$^{*}$\\[1mm]
$^{1}$Department of Communication Systems, KTH Royal Institute of Technology, Stockholm, Sweden\\
$^{2}$CMVS, University of Oulu, Finland\\
E-mail: yunquan@kth.se, chen.haoyu@oulu.fi\\[1mm]

}

\maketitle

\begin{abstract}
%  Multimodal Visual Language Models (VLMs), based on their powerful ability to understand images/videos, have been increasingly applied in various fields in recent years. In the research of AI's learning and understanding of animal and human behavior, VLMs play an important role. As compared to human, VLMs are able to observe more subtle features in animal and human behavior and make deeper understandings. Identifying leaders within a crowd is an important task for human society. However, there is a lack of research on using AI to find leaders in the crowd. To fill this gap, we replaced humans with mice for experiments, using currently the most advanced VLMs combined with unsupervised learning methods in deep learning to learn mouse behavior and predict the outcome of different mouse duels in narrow tube experiments. By comparing the predicted results of different VLMs with the actual outcome of mouse narrow tube experiments, we presents MNT-Bench (Mouse-Narrow-Tube-Benchmark), a novel
% mouse behavior benchmark that emphasizes the importance of different VLMs and parameters of unsupervised learning
%  for advanced mouse behavior understanding capability
% benchmarking. MNT-Bench evaluates the ability of ...under three distinct scenarios:

Understanding social dominance in animal behavior is critical for neuroscience and behavioral studies. In this work, we explore the capability of Multimodal Large Language Models (MLLMs) to analyze raw behavioral video of mice and predict their dominance hierarchy. We introduce MTT-Bench, a novel benchmark comprising annotated videos of pairwise mouse interactions for Mouse Tube Test analysis. Building on existing MLLM architectures, we fine-tune these models to perform zero-shot inference on unseen behavioral sequences, predicting social dominance without explicit labels during testing. Our framework demonstrates promising results, showing high agreement with tube test rankings. This work opens a new direction for applying foundation models to ethology and social behavior analysis, without the need to design domain-specific models.
\end{abstract}


\section{Introduction}
The study of mouse behavior has long been a central focus in bioinformatics and neuroscience-related computational research, primarily because mice serve as one of the most widely used model organisms for understanding human brain function and disease\cite{b1}. Their genetic, anatomical, and behavioral similarities to humans make them ideal subjects for translational studies\cite{b2}. Recent experiments have shown that the study of mouse behavior can also help people solve problems in the field of mental illness, Datta et al. emphasized that precise behavioral quantification in rodents provides a critical bridge between
neural activity and higher-order cognitive processes, which is essential for understanding psychiatric disorders\cite{b3}. 

With the rapid development of machine learning, many novel methods have emerged for studying mouse behavior. In particular, high-throughput behavioral phenotyping, combined with modern machine learning and bioinformatics tools, enables researchers to extract detailed insights into neuropsychiatric conditions, social interaction, stress response, and cognitive functions\cite{b4}. Recent advances in computer vision and deep learning have further expanded the capacity to automatically quantify subtle and complex mouse behaviors from video data, making behavioral analysis more objective, scalable, and reproducible. 

The study of mouse behavior has significant implications for investigating the relationship between humans and society. Among various behavioral paradigms used to investigate social interaction in mice, the mouse tube test has emerged as a widely adopted and robust method for assessing social dominance. In this test, two mice are introduced from opposite ends of a narrow tube that allows only one mouse to pass through at a time. The animal that forces its opponent to retreat is considered socially dominant\cite{b5}. With the integration of deep learning-based video analysis and unsupervised behavior classification, the tube test is increasingly used not only to evaluate social rank but also to explore individual traits such as submissiveness, aggression, and coping strategies under social stress\cite{b6}.

In this study, we conducted eleven groups of mouse narrow tube experiments, each group consisting of four training videos and one or two testing videos. Then we used the method of data augmentation to increase our data volume by eight times, in order to evaluate the performance of the MLLMs. We use state-of-the-art MLLMs to learn the characteristics of the mice in our experimental videos, allowing them to predict the results of the mouse narrow tube experiments. By comparing the predicted results of the MLLMs with the actual results, we introduce MTT-Bench (Mouse-Tube-Test-Benchmark) to evaluate MLLMs’ prediction capability of mouse narrow tube experiment results. 

Our main contributions are:
\begin{itemize}
 \item A new multimodal benchmark, MouseRank, linking naturalistic mouse interaction videos with tube-test-validated dominance labels.

 \item A fine-tuning framework for applying existing MLLMs to behavior-based dominance prediction, with zero-shot generalization capability.

 \item Empirical evidence that MLLMs can capture latent social traits in animal behavior, advancing the intersection of foundation models and ethology.
\end{itemize}

This work opens a novel perspective on how general-purpose multimodal models can be repurposed for scientific understanding of animal social cognition, pushing the boundaries of zero-shot reasoning in biological settings.

%Before deep learning was widely applied to the study of mouse behavior, Wiltschko et al. introduced a data-driven approach to identify sub-second behavioral motifs in freely behaving mice using unsupervised learning techniques, demonstrating the power of computational methods to reveal structure in spontaneous behavior without requiring predefined labels5. Unsupervised learning has since proven to be particularly well-suited for studying mouse behavior, largely because behavioral data—especially from continuous, naturalistic settings—are often high-dimensional, noisy, and lack clear annotation. These methods can discover latent structure in behavior without requiring manual labeling, which is both time-consuming and subjective. For example, Berman et al. applied unsupervised t-SNE embedding and Gaussian mixture modeling to track and categorize stereotyped Drosophila and mouse behaviors based purely on postural dynamics, establishing a framework that bypasses predefined ethograms6.

%\subsection{Maintaining the Integrity of the Specifications}

%The IEEEtran class file is used to format your paper and style the text. All margins, 
\section{Related Works}
\noindent\textbf{Automated Animal Behavior Analysis.} Quantifying animal behavior has long been a central goal in neuroscience and ethology. Traditional approaches often rely on manual annotations or simple tracking systems to extract locomotor or social interaction features\cite{b7}. Recent developments such as DeepLabCut and SLEAP have enabled markerless pose estimation in animals, facilitating large-scale behavioral analysis with improved spatial accuracy\cite{b8}\cite{b9}. However, while pose tracking has advanced, most downstream behavior classification pipelines still rely on domain-specific heuristics or supervised learning using pre-defined ethograms. Several works have proposed unsupervised or weakly supervised approaches to discover behavioral motifs from pose trajectories or video embeddings (e.g., MotionMapper, B-SOiD) \cite{b10}\cite{b11}. Yet, these methods typically require handcrafted feature engineering and fail to generalize across tasks such as social interaction or dominance hierarchy. Our work departs from this trend by directly leveraging vision-language pretraining from MLLMs to reason over complex, unstructured behavioral interactions.

\noindent\textbf{Social Dominance Inference in Mice.} Social dominance plays a crucial role in regulating access to resources, mating opportunities, and stress responses in group-housed animals. The tube test is one of the most widely used assays to measure pairwise dominance in rodents, where the mouse that pushes the other out of a narrow tube is considered dominant\cite{b12}. While widely accepted, this method does not capture naturalistic social behaviors and is time-consuming to scale. Recent work has attempted to predict dominance hierarchies using automated tracking and behavioral features, such as push events, following, or retreat behavior. For example, Zhou et al. (2022) proposed a spatiotemporal GCN to model mouse social interaction for dominance prediction\cite{b13}. However, such models rely heavily on pose accuracy and handcrafted behavioral priors. In contrast, we introduce a new paradigm where pre-trained MLLMs are fine-tuned to capture subtle behavioral dynamics and produce zero-shot dominance predictions from raw interaction videos.

\noindent\textbf{Multimodal Large Language Models for Scientific Reasoning.} Multimodal large language models (MLLMs), such as BLIP-2 \cite{b25}, OpenFlamingo \cite{b26}, and InternVL \cite{b27}, have shown remarkable success in visual reasoning, captioning, and video understanding tasks. By jointly modeling image/video inputs with textual prompts, these models can support few-shot or zero-shot generalization in open-ended domains. Several recent studies (e.g., MouseGPT, 2025) have explored using vision-language pretraining to analyze rodent behavior with natural language annotations\cite{b14}. However, to our knowledge, no prior work has applied MLLMs to predict latent traits such as social dominance from free-behaving animal interactions. Our work is among the first to investigate whether such general-purpose models can internalize complex ethological constructs like social hierarchy, even in the absence of direct supervision on test data.

\section{Materials and Methods}
%Before you begin to format your paper, first write and save the content as a 
%separate text file. Complete all content and organizational editing before 
%formatting. Please note sections \ref{AA} to \ref{FAT} below for more information on 
%proofreading, spelling and grammar.

%Keep your text and graphic files separate until after the text has been 
%formatted and styled. Do not number text heads---{\LaTeX} will do that 
\begin{figure*}[!t]
    \centering
    \includegraphics[width=\textwidth, height=0.55\textheight]{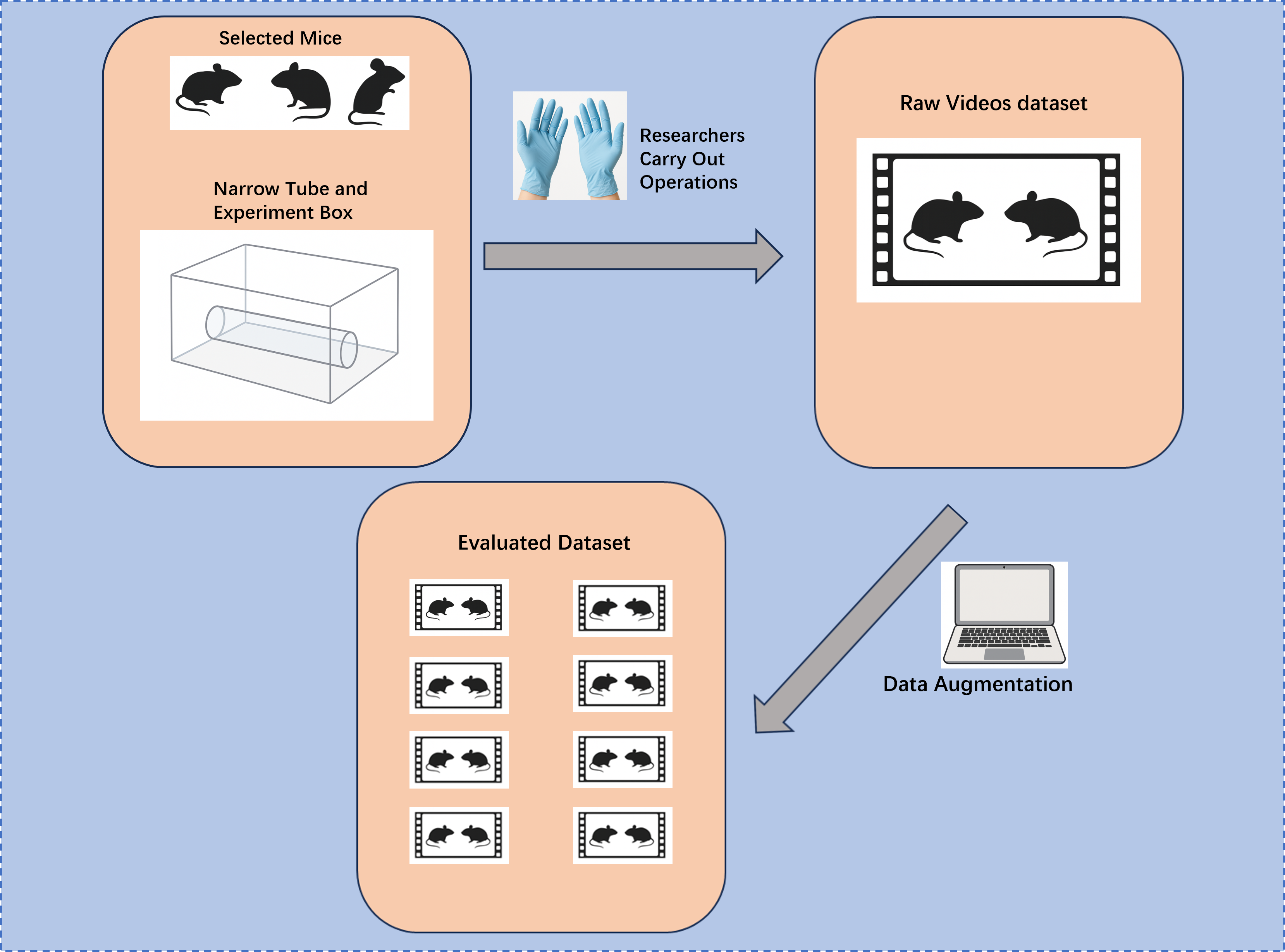}
    \caption{ Overview of the process of generating the Evaluated Dataset.}
    \label{fig:p1}
\end{figure*}

\subsection{Dataset}\label{AA}
For the purpose of evaluating the potential of explicitly leveraging state-of-the-art MLLMs for understanding and prediction of results in mouse tube test experiments, we collaborate with animal science experts specializing in studying mouse behavior to create a new dataset for mouse tube test experiments. We initially conducted thirteen experiments, each consisting of four training videos and one to two test videos of mice engaging in tube confrontation.
In each of the four training videos for each experiment, the two mice participating in this experiment each have two separate training videos. To record these experimental videos, researchers first allow mice to freely explore or walk through a transparent tube placed in their home cage for three consecutive days. And the video recording time of different groups is often separated by several days to allow the mice to fully adjust. The operations in these videos comply with the standard training procedure for mouse tube test experiments\cite{b15}, which requires each mouse to independently practice passing through the tube. The researchers will let the mouse pass multiple times a day (e.g. five times), and if the mouse stagnates, it can be gently pushed through to ensure learning the correct path.
In each group's testing video, we had two mice in the group engage in tube confrontation, the two mice enter from the left and right ends of the tube, and the dominant mouse ultimately pushes the other out of the tube, winning. The results of these testing videos will serve as the basis for evaluating the performance of the MLLMs.
However, due to the fact that in the two testing videos of the sixth experiment, two mice each won one tube confrontation, we were unable to determine the winner or loser relationship between the two mice. Therefore, we discarded the experimental data of this group. In the tenth experiment, the two mice showed abnormal behaviors, which we believed was due to external interference, so we also discarded the data of this group. In general, we obtained data from eleven groups of mouse narrow tube experiments, including 44 training videos and 14 testing videos, which was our initial dataset.

Due to the large amount of experimental data required to evaluate the performance of the MLLMs, we performed data augmentation on the initial dataset. There are many methods of data augmentation that have been proven useful in deep learning\cite{b16}, in this dataset, we choose four common methods for data augmentation: 1.flip: Flip the video so that its left and right sides are swapped;
2.gray1: Adjust the video grayscale to increase its brightness; 3.gray2: Adjust the video grayscale to reduce its brightness; 4.rotated: Rotate the video clockwise by 15 degrees. Due to the fact that a good way to augment data is to mix different data augmentation methods\cite{b17}, we ultimately expanded the original dataset by eight times: original, flip, gray1, gray2, gray1-flip, gray2-flip, rotated-original, rotated-flip. Therefore, we obtained a dataset containing 88 groups of mouse narrow tube experiments (each contains of four training videos) to evaluate the performance of the MLLMs. The process of generating the Evaluated Dataset is shown in Fig .1.

\begin{figure*}[!t]
    \centering
    \includegraphics[width=\textwidth, height=0.8\textheight]{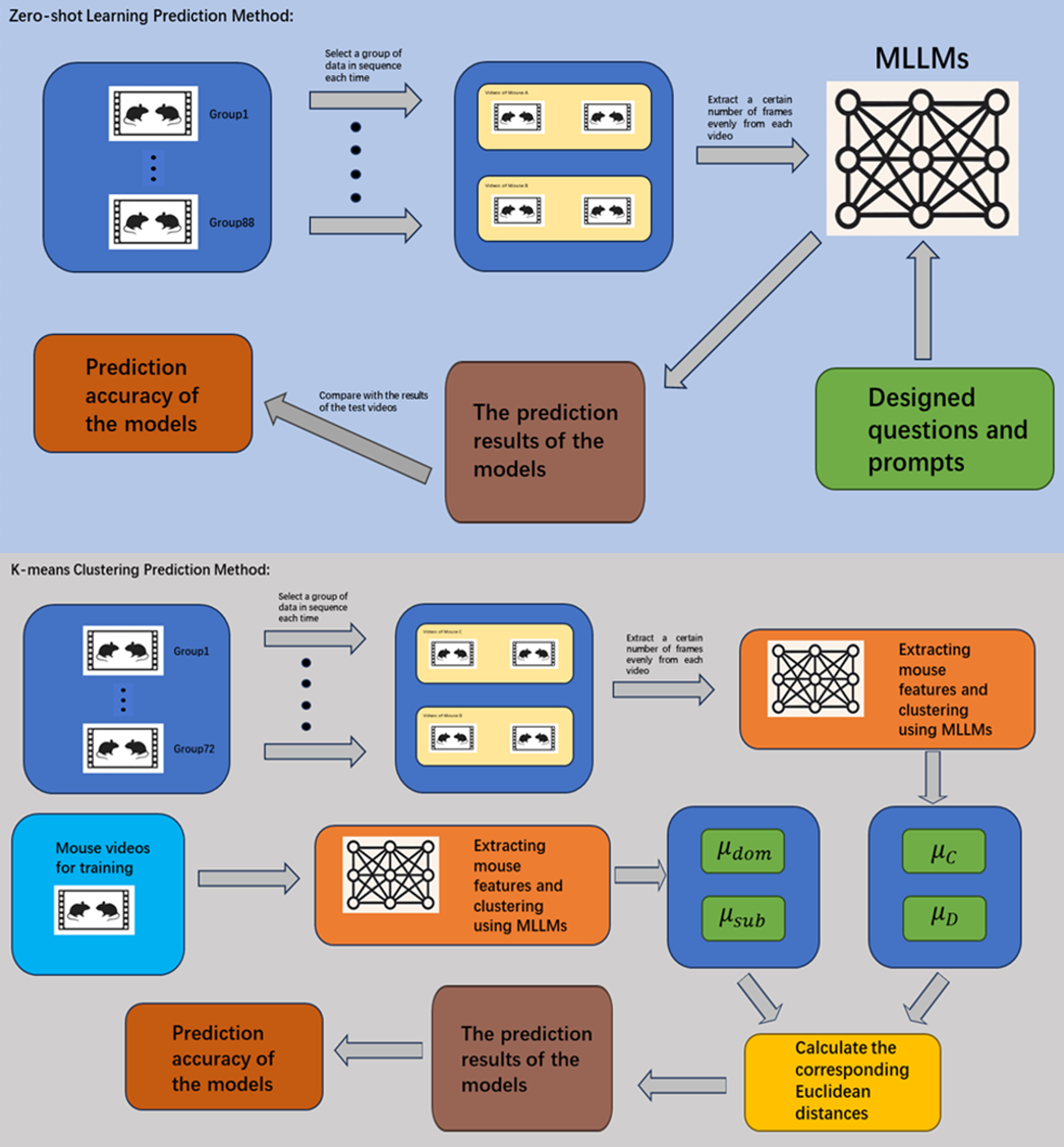}
    \caption{ The operation process of the two prediction methods.}
    \label{fig:p2}
\end{figure*}

\subsection*{B. \textit{Propose Method}}
In this section, we present the construction process of our MTT-Bench. We start with a detailed introduction to the two different methods of using MLLMs to predict the results of mouse tube test experiments, and then comprehensively described the implementation process from the dataset to the predicted results. At the end of this section, we described the experimental method for artificially predicting the results of the mouse tube test experiments, because the results of human prediction can be compared with the MLLMs' prediction.
\begin{enumerate}[label=\arabic*), align=left, leftmargin=0pt, labelsep=0.5em, itemindent=1.5em, listparindent=0pt]
    \item \textbf{\textit{Zero-shot Learning Prediction Method:}} %In this study, we introduce a new deep learning approach BGCCGB to predict BGCs in genomes and classify their product types, as shown in Fig. 1. 
    Zero-shot learning (ZSL) is a common method in MLLMs' video understanding tasks. Zero-shot learning (ZSL) enables a model to recognize classes or perform tasks it has never seen during training by leveraging semantic information such as textual descriptions. This is particularly useful in domains where labeled data for all classes is hard to obtain. 
    In formal terms, ZSL assumes a training set with samples 
$\mathcal{D}_{\text{train}} = \{(x_i, y_i)\}_{i=1}^N$, 
where $y_i \in \mathcal{Y}_{\text{seen}}$, and a test set 
$\mathcal{D}_{\text{test}} = \{(x_j)\}_{j=1}^M$, 
where the labels $y_j \in \mathcal{Y}_{\text{unseen}}$, and 
$\mathcal{Y}_{\text{seen}} \cap \mathcal{Y}_{\text{unseen}} = \emptyset$. 
ZSL leverages semantic embeddings of classes (e.g., text prompts) 
$\phi(y) \in \mathbb{R}^d$, and visual encodings 
$\psi(x) \in \mathbb{R}^d$, and predicts via similarity:
\begin{equation}
\hat{y} = \arg\max_{y \in \mathcal{Y}_{\text{unseen}}} \, \text{sim}(\psi(x), \phi(y))
\end{equation}
where $\text{sim}(a, b)$ is often cosine similarity. This formulation allows generalization to unseen classes through shared latent space alignment between vision and language. Recent studies such as CLIP\cite{b18}, Flamingo\cite{b19}, and VideoCoCa\cite{b20} demonstrate the effectiveness of vision-language pretraining for zero-shot tasks. These models are capable of performing well across multiple domains with only task-descriptive text input, and without requiring fine-tuning on specific datasets.
In this study, the dataset does not have labels for mouse behavior, and it is difficult to label the behavior of mice in the training videos in order to achieve the goal of predicting mouse narrow tube experiments. Therefore, Zero-shot learning (ZSL) is highly applicable to this study. In the context of mouse behavior analysis, MLLMs such as InternVL can process frames from mouse videos and match them against natural language prompts describing behavioral traits, such as “a dominant mouse pushing forward” or “a submissive mouse backing away.” This avoids the need for explicit behavior labeling while enabling interpretable predictions. The prediction process is shown in Fig .2. In each group of experimental data, we named the two mice participating in the experiment as Mouse A and Mouse B. The mouse with the smaller label is Mouse A, and the one with the larger label is Mouse B (for example, Mouse 4 is Mouse A, and Mouse 5 is Mouse B). Then we use MLLMs to read the content of these four training videos and learn the behavioral characteristics of the mice. To predict the results of the narrow tube experiment, we ask MLLMs three questions and provide prompts in the questions:

Question1: We provide the behavioral parameters of mouse A obtained through code, and then ask the model: "Output the personality score of the mouse between 0 and 1, with the strongest personality scoring 1 and the weakest personality scoring 0."

Question2: Similar to question 1, we provide the behavioral parameters of mouse B and then ask the model: "Output the personality score of the mouse between 0 and 1, with the strongest personality scoring 1 and the weakest personality scoring 0."

Question3: Let the model synthesize the answers of the first two questions and then ask: "Assuming that a higher score means a more dominant personality," ; "Choose one of these two mice, who would win in a narrow tube test and why?"

Based on the model's answers to these three questions, the prediction results of MLLMs for each group of mouse narrow tube experiments can be clearly obtained. Then, these prediction results of all the 88 groups can be compared with the actual results of mouse tube test experiments showed the testing videos, and multiple predictions can be made to take the average to obtain the prediction accuracy of the MLLMs. This is the prediction accuracy obtained by combining MLLMs with Zero-shot Learning (ZSL).

\vspace{0.2\baselineskip}
    \item \textbf{\textit{K-means Clustering Prediction Method:}} %The efficacy of deep learning models is heavily dependent on the quantity of training data. However, the MIBiG database suffers from a limited number of overall BGC records, as well as unbalanced distribution across specific BGC categories (e.g., only 39 Alkaloid records). This issue of limited data quantity has not been adequately addressed in prior research on this topic.
K-means clustering is a classic unsupervised learning method that demonstrates a certain ability in understanding videos and can identify potential structures in data without the need for labels \cite{b21}. Although K-means clustering may not have the same understanding ability for videos as some other methods, it still has certain reference value. In this study, we introduced a prediction method based on K-means and combined it with the MLLMs to extract behavioral features of mice and predict the results of mouse tube test experiments. This method enables interpretable and unsupervised prediction of mouse dominance from behavioral video data without the need for dense labels. The prediction process is shown in Figure .2. Due to the difficulty of GPUs in withstanding excessively large training volumes, we only selected the training videos of the first five groups in the original dataset to train the model. Among these 20 training videos, 10 of the videos are about mice won narrow tube experiment, we classified these 10 videos into one category, which is category1; And in the other 10 videos, the mice lost the narrow tube experiment. We classified these 10 videos into one category, which is category2. The number of clusters in the K-means clustering method is set to 4 because when using the trained model to predict the results of other groups, each group has four training videos, and the number of clusters must be less than or equal to the number of training videos of each group. After setting these parameters, we first use the pretrained MLLMs to encode each video of category1 $v \in \mathcal{V}_1$ into a high-dimensional embedding: \begin{equation}
z = \psi(v), \quad z \in \mathbb{R}^d,
\end{equation}
where $\psi(\cdot)$ is the pretrained MLLMs' encoder and $d$ is the embedding dimension. We also perform the same operation on category2. 
We then aggregate the embeddings of these two categories respectively to compute two feature centers via:
\begin{equation}
\mu_\text{dom} = \frac{1}{|\mathcal{V}_\text{1}|} \sum_{v \in \mathcal{V}_\text{1}} \psi(v), \quad
\mu_\text{sub} = \frac{1}{|\mathcal{V}_\text{2}|} \sum_{v \in \mathcal{V}_\text{2}} \psi(v).
\end{equation}
where \(\mathcal{V}_1\) represents category1 and \(\mathcal{V}_2\) represents category2.
These two feature centers, $\mu_\text{dom}$ and $\mu_\text{sub}$, belonging to category1 and category2 respectively, serve as the behavioral prototypes for dominant and submissive mice, respectively. Similar embedding-based aggregation approaches have been applied in animal behavior clustering \cite{b22}, self-supervised learning \cite{b23}, and video understanding \cite{b24}.

Due to the selection of these five groups of data as the training set, we chose the remained six groups of data from the original dataset as the testing set, as well as the videos obtained by data augmentation on these six groups of data. However, even so, the testing set only contained 48 groups, so we further enhanced the original six groups of data, four new data augmentation methods except the eight methods mentioned in the second paragraph of III.A were used : ro-g1 (rotated-gray1), ro-g2 (rotated-gray2), ro-g1flip (rotate-gray1-flip) and ro-g2flip (rotate-gray2-flip). Therefore, we have 72 groups of data as the testing set. When predicting the results of a tube test experiment, we used the four training videos of this group and calculated the feature centers of the behavior of the two mice in this group using the method mentioned above. Assuming these two mice are mouse C and mouse D, calculate the Euclidean distances \(d1\) and \(d2\) from the feature center of mouse C to \(\mu_{\text{dom}}\)
and \(\mu_{\text{dom}}\).
Similarly, the relevant parameters of mouse D are \(d3\) and \(d4\). The formula for calculating the Euclidean distance between two feature centers in K-means clustering is as follows:
\begin{equation}
D(C_i, C_j) = \left\| C_i - C_j \right\|_2 = \sqrt{ \sum_{k=1}^{d} (C_i^{(k)} - C_j^{(k)})^2 }
\end{equation}
\noindent
where \( C_i \) and \( C_j \) are two feature centers (i.e., cluster centroids), \( d \) is the embedding dimension, \( C_i^{(k)} \) denotes the \( k \)-th component of centroid \( C_i \), and \( \left\| \cdot \right\|_2 \) denotes the Euclidean norm.

In K-means clustering, the Euclidean distance between two feature centers can reflect the degree of similarity between two individuals to a certain extent. The smaller the distance, the more similar they are, and the larger the distance, the less similar they are. Therefore, we first compare \(d1\) and \(d2\), if \( d1>d2 \), mouse C belongs to advantage class, else, it belongs to disadvantage class, mouse D is the same. If one mouse belongs to  advantage class and the other belongs to disadvantage class, it is judged that the mouse belongs to advantage class wins. If they all belong to
advantage class, if \( d1<d3 \), mouse C wins, else, mouse D wins. If they all belong to disadvantage class, if \( d2>d4 \), mouse C wins, else, mouse D wins. Finally, the obtained prediction results are processed in the same way as the prediction results of Zero-shot Learning Prediction Method to obtain the accuracy of K-means Clustering Prediction Method.
  %  \item 
%\vspace{0.2\baselineskip}
%\textbf{\textit{Fine-tuning:}} 

    \item
\vspace{0.2\baselineskip}
    \textbf{\textit{Human Agent:}} 
In order to obtain an artificial prediction accuracy that has reference value for the accuracy of model predictions, when using ”Human Agent” prediction method, we searched for 30 individuals from various age groups and social backgrounds
who had no prior knowledge of our dataset. After watching the training videos of these 88 groups of mice, they were asked
to make judgments on the outcome of each group of mice. Then, we took the average accuracy of everyone’s responses.
\end{enumerate}

\begin{table*}[ht]
\centering
%\caption{Detailed evaluation results on MNT-Bench.}
\begin{adjustbox}{width=\textwidth}
%\begin{tabular}{llccccc|c}
\renewcommand{\arraystretch}{1.3}
\begin{tabular}{c|c|ccccc|c}
%\toprule
%\toprule
\hline
\hline
\textbf{Model} & \#Frames & Prediction1 & Prediction2 & Prediction3 & Prediction4 & Prediction5 & \textbf{Overall Avg.} \\
%\midrule
\hline
\multicolumn{8}{c}{\textbf{Human}} \\ 
%\midrule
\hline
Human Agents & - & - & - & - & - & - & 51.44 \\
%\midrule
\hline
\multicolumn{8}{c}{\textbf{Zero-shot Learning Prediction Method}} \\
%\midrule
\hline
InternVL3-1B & 32 & 73.86 & 64.77 & 75.00 & 78.41 & 69.32 & 72.27  \\
%\midrule
InternVL3-2B & 64 & 77.27 & 78.41 & 72.73 & 77.27 & 75.00 & 76.14  \\
InternVL3-8B & 64 & 63.64 & 68.18 & 64.77 & 72.73 & 65.91 & 67.05  \\
BLIP-2\_t5 & 32 & 37.50 & 39.77 & 44.32 & 48.86 & 45.45 & 44.32 \\
GPT-4o & 32 & 15.91 & 19.32 & 23.86 & 22.73 & 25.00 & 21.36  \\
GPT-4-turbo & - & 18.18 & 17.05 & 20.45 & 23.86 & 26.14 & 21.14  \\
Gemini-1.5-Pro & 0.2fps & 44.32 & 50.00 & 51.14 & 56.82 & 53.41 & 51.14  \\
\hline
\multicolumn{8}{c}{\textbf{K-means Clustering Prediction Method}} \\
%\midrule
\hline
InternVL3-1B & 256 & 56.94 & 50.00 & 59.72 & 52.78 & 51.39 & 54.17  \\
%\midrule
%\hline
%\multicolumn{8}{c}%{\textbf{Fine-tuning}} \\
%\midrule
%\hline
%InternVL3-8B & 64 & - & - & - & - & - & - \\
\hline
\hline
%\bottomrule
%\bottomrule
\end{tabular}
\end{adjustbox}
\caption{Detailed evaluation results on MTT-Bench.}
\label{tab:mnt}
\end{table*}
\section{Experiments}
This section presents comprehensive experiments and indepth analyses of MTT-Bench.
\subsection{Models and Evaluation Strategies}
We evaluate three existing types of models: (1) Ofﬂine
Multimodal Models, including BLIP-2\cite{b25}, GPT-4o \cite{b28}, InternVL3-1B \cite{b29}, InternVL3-2B \cite{b30}, InternVL3-8B \cite{b31} and Gemini-1.5-Pro \cite{b32} (2) Blind LLMs, including GPT-4-turbo \cite{b33} (3) Human Agents. When using the same prediction method, to ensure a fair comparison of model performance, we adhere as much as possible to the principle of consistency by maintaining the same number of frames or frames per second (fps) across all models. When different prediction methods are used, due to the different characteristics of the prediction methods themselves, the number of frames or frames per second (fps) of the models might be different.

\subsection{Main Results
}
Table~\ref{tab:mnt} reports the performance of seven models on MTT-Bench. When combining an MLLM with a certain prediction method, we conducted five times of predictions, and the results of these five predictions and their average were used as the basis for evaluating the performance of this combination. Our evaluation brings several important findings, as follows:

\textbf{We used the zero-shot learning prediction method in seven models. Among these models, InternVL3-1B, InternVL3-2B and InternVL3-8B belong to the InternVL series.} Among the three, InternVL3-1B, InternVL3-2B, and InternVL3-8B differ mainly in scale and capability. InternVL3-1B is lightweight and efficient, while InternVL3-2B and 8B offer progressively stronger performance with increased model size and computational cost. The model capacity of InternVL3-1B is relatively small, and the maximum input token limit is low, so when we combined this model with zero-shot, we extracted 32 frames from each video. On the other two models, we chose 64 frames per video. In terms of average accuracy, InternVL3-2B performs the best at 76.14\%, followed by InternVL3-8B at 72.27\%, and InternVL3-1B performs the worst at 67.05\%. In addition, the reason why InternVL3-1B made prediction errors in many groups was that the model could not obtain the behavioral scores of the mice in question 1 and question 2, so the results could not be predicted. We believe this is because InternVL3-1B has insufficient computing power when handling such tasks. In the prediction results of InternVL3-8B, there are also many groups whose prediction results cannot be predicted, and the number is even more than that in InternVL3-1B. But at this point, the prompt information output by the model is usually like this :"Given that both Mouse A and Mouse B have the same personality score of 0.7, it's difficult to definitively predict which one would win. Both mice are likely to perform similarly in the test. If there are other factors such as size, strength, or experience, those could also influence the outcome." We argue that this does not indicate model degradation, but rather that the larger model becomes more “cautious” or “conservative” when the input features are ambiguous or when behavioral differences are subtle. Specifically, InternVL3-8B tends to generate uncertain or neutral responses—such as "both mice have the same personality score, making it difficult to determine a winner"—when it cannot extract clearly differentiable behavioral cues from the input. In contrast, the smaller InternVL3-1B model, despite its limited capacity, is more likely to make “bold” predictions based on shallow patterns, which may result in higher apparent discriminability under the same ambiguous conditions. In order to fully utilize the capabilities of InternVL3-8B, we plan to combine it with the Fine-tuning method in our future research.
The situation of InternVL3-2B is much better, usually only results of one or two groups can't be predicted, and the accuracy of this model is also the highest among the three.

\textbf{The performance of BLIP-2 is worse than InternVL series models on this task.} As a MLLM launched earlier than InternVL, BLIP-2 shares many similarities with the models of the InternVL series. Due to weaker inference efficiency compared to InternVL, the prediction accuracy of BLIP-2 is lower. Although the model can provide clear answers in predictions of each group, the final prediction accuracy is only 44.32\%, which is even lower than the human agent prediction accuracy 51.44\%.

\textbf{Also combined with zero-shot, the performance of GPT-4o and GPT-4-turbo is not optimistic.} GPT-4o is an offline Multimodal Models, but due to its limited computing capacity, we took 32 frames per video and the overall accuracy of this model is only 21.36\%. We also changed our way of asking questions in order to make the output of the model match our goals. We initially attempted to ask the model the three questions mentioned in III-B-1), but the answer of GPT-4o is :"I'm unable to view or analyze video content or images to determine a personality score for the mouse. Personality scores are usually based on specific behavioral assessments conducted by researchers or experts in the field. If you have a specific methodology or criteria for scoring personality, you might apply that based on your observation of the video content." So we changed our questions and asked the model :"Please give this mouse a personalized rating based on the following criteria:
- Activity level: whether it often move or run;
- Resistance desire: strong resistance to the interference of researchers;
- Exploration desire: whether it often explore the environment;
Please score 1-5 points for each item and calculate the total score.
Here are its video frames."
After analyzing all the problems encountered during the prediction process and the predicted output results, we found that the main reason for the low prediction accuracy of GPT-4o is that the prediction results of this model of about half of the groups are unable to predict the outcome of mice tube test, and more information is needed. For example, in Prediction 1, the number of unpredictable groups is 45, the number of correctly predicted groups is 14, and the number of incorrectly predicted groups is 29. The performance of GPT-4-turbo is similar to that of GPT-4o, but as a Blind LLM, GPT-4-turbo performed worse on this task. In general, GPT series models perform poorly in handling this task.
And due to lack of permission, we are also unable to combine GPT series models with the Fine-tuning method.

\textbf{The predictive performance of Gemini-1.5-Pro is also quite distinctive.} Inspired by the OVO-Bench production process \cite{b34}, we took one frame every five seconds from the training videos and fed it into the Gemini-1.5-Pro model. Unlike other models combined with zero-shot method, Gemini-1.5-Pro provided clear prediction results for all 88 groups of mice in the narrow tube experiment. Ultimately, the average prediction accuracy of Gemini-1.5-Pro is 51.14\%, slightly lower than the human agent prediction accuracy 51.44\%.

\textbf{We used K-means clustering prediction method only on InternVL3-1B.} That's because although this method seems to be designed very reasonably, but K-means have limitations: it is an unsupervised learning algorithm that clusters data based on similarity in feature space without any semantic understanding or task-specific supervision. Consequently, it cannot reason about high-level behavioral traits or infer subtle contextual cues necessary for accurate outcome prediction. In our task, which involves assessing social dominance in mice based on behavioral video data, these limitations become critical. The prediction relies not only on visual similarity but also on nuanced temporal dynamics and behavioral interpretation—something K-means is fundamentally incapable of capturing. Therefore, while the K-means clustering prediction method provides a baseline, its performance is significantly lower than that of language-guided models combined with the zero-shot learning prediction method. However, K-means clustering prediction method has some advantages in video processing. We found in the experiment that compared to the zero-shot learning prediction method, the K-means clustering prediction method demonstrates a higher frame-processing capacity for video data. Specifically, due to the input token length limitation of the zero-shot method, each video is typically restricted to 64 frames, and in the case of InternVL3-1B, only 32 frames per video are used. In contrast, the K-means method relies solely on pre-extracted visual feature embeddings, which are more compact and memory-efficient, allowing the number of frames per video to scale up to 256. Moreover, the K-means method offers greater flexibility in data augmentation. Since it does not depend on sequential or token-based input structures, it can incorporate various visual transformations, temporal cropping, and ensemble averaging across multiple video segments. These advantages make it a practical choice in large-scale data processing scenarios or environments with limited computational resources. In our experiments combining the K-means clustering prediction method with InternVL3-1B, we used 256 frames per video and applied 11 types of data augmentation. We also attempted to incorporate the four newly introduced augmentation strategies from the K-means clustering into the zero-shot learning experiments, but found that some models failed to run successfully. As a result, we ultimately excluded these augmentations from the zero-shot learning experiments. Nevertheless, K-means remains incapable of modeling contextual information, behavioral intent, or nuanced behavioral interpretation, which limits its performance in our task of fine-grained social interaction discrimination. Under this method, the final average prediction accuracy was 54.17\%, slightly higher than the human agent prediction accuracy of 51.44\%, but significantly lower than the accuracy achieved by the zero-shot learning prediction method.

\section{Discussion}
In this work, we introduced MTT-Bench, a comprehensive benchmark designed to assess mouse tube test experiment result prediction capabilities of MLLMs across two prediction methods: zero-shot learning and K-means clustering. We anticipate that MTT-Bench will serve as a valuable resource for the research community, guiding the development of MLLMs toward the field of animal behavior understanding and prediction. We also hope to encourage research on multimodal models in high-level social understanding by establishing clear evaluation protocols and opening benchmark data and tools. In the future, we hope to expand MTT-Bench to more new MLLMs and add Fine-tuning as a prediction method, paving the way toward generalized animal behavior intelligence in AI systems.

%\section*{Acknowledgment}

%\section*{References}

%Please number citations consecutively within brackets \cite{b1}. The 
%sentence punctuation follows the bracket \cite{b2}. Refer simply to the reference 
%number, as in \cite{b3}---do not use ``Ref. \cite{b3}'' or ``reference \cite{b3}'' except at 
%the beginning of a sentence: ``Reference \cite{b3} was the first $\ldots$''

%Number footnotes separately in superscripts. Place the actual footnote at 
%the bottom of the column in which it was cited. Do not put footnotes in the 
%abstract or reference list. Use letters for table footnotes.

%Unless there are six authors or more give all authors' names; do not use 
%``et al.''. Papers that have not been published, even if they have been 
%submitted for publication, should be cited as ``unpublished'' \cite{b4}. Papers 
%that have been accepted for publication should be cited as ``in press'' \cite{b5}. 
%Capitalize only the first word in a paper title, except for proper nouns and 
%element symbols.

%For papers published in translation journals, please give the English 
%citation first, followed by the original foreign-language citation \cite{b6}.

\end{document}